# Design optimization of advanced tow-steered composites with manufacturing constraints

Chuan Luo[a,*], Federico Ferrari[a,b], James K. Guest[a]

[a]*Civil and Systems Engineering Department, Johns Hopkins University, 3400 N. Charles Street, Baltimore, MD, 21218, USA*
[b]*Department of Civil and Mechanical Engineering, Technical University of Denmark, Koppels Allé 404, 2800 Lyngby, Denmark*

**Abstract**

Tow steering technologies, such as Automated fiber placement, enable the fabrication of composite laminates with curvilinear fiber, tow, or tape paths. Designers may therefore tailor tow orientations locally according to the expected local stress state within a structure, such that strong and stiff orientations of the tow are (for example) optimized to provide maximal mechanical benefit. Tow path optimization can be an effective tool in automating this design process, yet has a tendency to create complex designs that may be challenging to manufacture. In the context of tow steering, these complexities can manifest in defects such as tow wrinkling, gaps, overlaps. In this work, we implement manufacturing constraints within the tow path optimization formulation to restrict the minimum tow turning radius and the maximum density of gaps between and overlaps of tows. This is achieved by bounding the local value of the curl and divergence of the vector field associated with the tow orientations. The resulting local constraints are effectively enforced in the optimization framework through the Augmented Lagrangian method. The resulting optimization methodology is demonstrated by designing 2D and 3D structures with optimized tow orientation paths that maximize stiffness (minimize compliance) considering various levels of manufacturing restrictions. The optimized tow paths are shown to be structurally efficient and to respect imposed manufacturing constraints. As expected, the more geometrical complexity that can be achieved by the feedstock tow and placement technology, the higher the stiffness of the resulting optimized design.

*Keywords:* Tow steering; Composite laminates; Optimization; Manufacturing constraints

# 1. Introduction

Automated Fiber Placement (AFP) is an innovative manufacturing technology capable of laying fiber, yarn, or tow strips of a specified width along curvilinear paths over complex 3D surfaces, thus allowing the local "steering" of tows within each layer of a laminate. The resulting Variable Stiffness Laminates (VSL) feature curved tow paths that allow a continuous variation of structural properties within the laminate, often enabling improved structural performance compared to laminates composed of stacked plies with discrete fiber orientation [1–3].

* Corresponding author.
*E-mail address:* cl4486@columbia.edu

Continuously varying tow orientations expand the design space for composite laminate optimization. In very select cases, optimization can be achieved in closed form, while numerical optimization is leveraged for more general design cases. For example, the closed form solution for the fiber paths of an orthotropic laminate with maximum stiffness is given by Pedersen [4,5], as a function of the local ratio of two principal strains and nondimensional material parameters. However, these closed form solutions are limited to very specific instances involving single-loaded minimum compliance design and simple geometries. Several research works have shown the benefits of optimized tow-steered laminates, as opposed to classical ones, for structural applications, such as buckling capacity [6,7], stiffness [8–11], maximum fundamental frequency [12], and strength [13–15]. These more challenging design problems are solved using non-linear programming methods.

The most popular gradient-based approach to VSL optimization is the so-called Continuous Fiber Angle Optimization (CFAO) [16]. In CFAO, the design variables either describe the fiber orientations directly, by the local angles (so-called Polar representation [17]), or indirectly, by the associated vector field (so-called Cartesian representation [18]). The latter has proven useful to avoid low quality local minima, which are present in the problem due to non-convexity of energy functionals with respect to orientations [19–22]. Recently, a tensor representation of fiber orientations, coupled with multi-variable projection methods to satisfy tensor invariants constraints was proposed as an extension of the Cartesian approach [23]. We note these methods refer to "fiber" orientations, but are also applicable to tow-steering.

Although designs achieved by CFAO can show impressive structural improvements, they can be challenging to manufacture. Optimized VSL designs may contain orientation discontinuities in regions where the principal stresses have nearly equivalent magnitudes but opposite signs, as well as in low strain energy regions [24,25]. Designs with large variations of fiber orientations are practically hard to produce, calling for post-processing operations on the orientations that may destroy the optimized performance. To avoid such large variations, and to control the smoothness of the fiber paths, filtering of the orientation design variables has been introduced in CFAO [26–28]. However, orientation filtering usually requires a rather large filter radius to achieve fiber continuity, with a significant reduction in performance [18,23]. Moreover, orientation filtering does not imply that practical manufacturing constraints are fulfilled [29], as will be shown herein.

Other than the basic continuity of tows paths, the production of VSL by the AFP process demands specific manufacturing constraints, generally more restrictive than those for composites with unidirectional fibers. The most common constraint is that on the minimum allowable turning radius of the tow path. When a tow is steered, the different inner and outer radii generate a transition between compression and tension at its edges. This may lead to defects such as fiber buckling and wrinkling, that are likely to occur in tows with small steering radius [29]. The minimum value of steering radius to avoid defects mostly depends on material properties and the manufacturing equipment [30]. Another important limitation is the prevention of tow gaps and overlaps. Indeed, thickness build-ups occur in the regions with overlapping tows [31], leading to unwanted thickness variations of the laminate. Most importantly, tows gaps and overlaps introduce stress discontinuities and concentrations, likely causing detrimental interlaminar stresses and delamination failures.

To date, only a few works have been focused on the introduction of these manufacturing constraints in CFAO. Van Campen *et al.* [6], followed by others [32,33], achieved VSL with manufacturable orientations fulfilling minimum turning radius using a multi-step optimization, coupled with the indirect representation of orientation by the lamination parameters [34]. However, this approach requires additional post-processing to retrieve the physical fiber orientations matching the optimized lamination parameters and steering constraints. A novel orientation filtering technique was recently proposed by Jantos *et al.* [35], and the authors quantitatively linked the smoothing radius to tows curvature.

Tow and fiber gaps, overlaps, and curvature were first related to differential operators (*i.e.*, divergence and curl) of the underlying orientation field parameterized by B-splines [36]. Manufacturing constraints were also considered in works using alternate parametrizations for the tow/fiber paths, such as level set functions [37,38] and B-splines [39]. Despite reducing the number of design variables, such methods offer less design freedom compared to discretized orientation design variables [30].

A common approach to deal with local constraints is to use aggregation functions, such as the *p*-norm (or *p*-mean), or the Kreisselmeier–Steinhauser (KS) function [40], to replace the many local constraints with a single aggregated constraint (or a few, based on spatial clusters). The approximation generally depends on the number of local constraints aggregated, their relative magnitude [41], and on the value of an aggregation (penalty-like) parameter. To keep a good approximation as the number of aggregated constraints increases, the aggregation parameter generally needs to be increased, increasing the nonlinearity of the problem, and potentially leading to ill-conditioning. Even if some strategies can be used to curb the growth of the aggregation parameter, such as normalizing the aggregation [42], apply continuation [43,44], or restricting the maximum change of design variables [41], aggregated constraints tend to reduce design freedom in the initial optimization stages and slow down convergence [44]. The Augmented Lagrangian (AL) function [45,46] can be used as an alternative to aggregation constraint approaches. The AL has been successfully applied to topology optimization problems with up to hundreds of millions local stress constraints [44,47–55], as well as design problems with many local volume constraints [56].

In this work, we investigate the effect of pointwise curvatures and gap/overlap control on the tow/fiber orientation field. These constraints are introduced in the optimization problem by bounding the maximum magnitude of the curl and divergence of the vector field representing the orientations, as proposed in [36]. In this way, the manufacturing constraints are directly linked to the physical orientations and not to derived fields, such as lamination parameters, thus avoiding the need for a reconstruction step. We then use the AL method to account for the large number of local constraints in the dual optimization algorithm, avoiding constraint aggregation strategies. The whole approach is shown to be algorithmically simple, and to provide qualitatively and quantitatively better designs compared to those achieved with orientation filtering methods.

The remainder of the manuscript is organized as follows. In Section 2, we review the orthotropic material law and orientation representation adopted in this work. Section 3 discusses the considered manufacturing constraints of minimum turning radius and gap/overlap control, and their link with the curl and divergence of the orientation field. In Section 4 we formulate a maximum stiffness (minimum compliance) optimization design problem including local manufacturability constraints and discuss the AL approach for their treatment. Three optimization examples are included in Section 5 to demonstrate the effectiveness of the proposed approach. Concluding remarks are provided in Section 6.

## 2. Anisotropic constitutive law and orientation representation

Here we describe the constitutive law used for the 2D plate examples of Sections 5.1 and 5.2, whereas we refer to Appendix A for the extension to the model used for the 3D shell example of Section 5.3.

We assume a homogeneous orthotropic material law for the composite, and linear elastic behavior under the applied loads. In the local material frame $\mathbf{X_0}$, defined by the principal material directions (see Fig. 1(a)), the stresses $\boldsymbol{\sigma_0} = \{\sigma_{11}, \sigma_{22}, \sigma_{12}\}$ and strains $\boldsymbol{\varepsilon_0} = \{\varepsilon_{11}, \varepsilon_{22}, 2\varepsilon_{12}\}$ are linked through $\boldsymbol{\sigma_0} = \boldsymbol{C_0}\boldsymbol{\varepsilon_0}$ where, under plane stress assumptions [57]

$$\boldsymbol{C_0} = \frac{1}{(1-\nu_{12}\nu_{21})}\begin{bmatrix} E_1 & \nu_{12}E_2 & 0 \\ \nu_{21}E_1 & E_2 & 0 \\ 0 & 0 & G_{12}(1-\nu_{12}\nu_{21}) \end{bmatrix} \quad (1)$$

where $E_1$, $E_2$ are the Young's moduli, $G_{12}$ is the shear modulus, and $\nu_{12}$ and $\nu_{21}$ are the Poisson's ratios.

Strains and stresses are then transformed to the reference coordinate system **X** (see Fig. 1(a)) as $\boldsymbol{\varepsilon}_X = \boldsymbol{Q}(\theta)^{-T}\boldsymbol{\varepsilon}_0$ and $\boldsymbol{\sigma}_X = \boldsymbol{Q}(\theta)^{-1}\boldsymbol{\sigma}_0$, through the matrix

$$\boldsymbol{Q}(\theta) = \begin{bmatrix} \cos^2\theta & \sin^2\theta & 2\cos\theta\sin\theta \\ \sin^2\theta & \cos^2\theta & -2\cos\theta\sin\theta \\ -\cos\theta\sin\theta & \cos\theta\sin\theta & \cos^2\theta - \sin^2\theta \end{bmatrix} \tag{2}$$

The constitutive tensor in the reference coordinate system is thus

$$\boldsymbol{C}_X = \boldsymbol{Q}(\theta)^{-1}\boldsymbol{C}_0\boldsymbol{Q}(\theta)^{-T} \tag{3}$$

In the Polar representation [58], the orientation field is described by assigning to each element centroid ($X_e$) one design variable, $\theta_e = \theta(X_e)$, representing the local orientation (see Fig. 1(b), left). While this choice requires only one design variable for each design point, it introduces challenges for gradient-based optimizers. First, due to the existence of multiple local minima in the strain energy density as a function of the orientations $\theta$, it is very likely to end up in a sub-optimal solution [4,20,59]. Second, the periodic nature of $\theta$ leads to a mathematical discontinuity, often referred to as "$2\pi$ ambiguity" [18,23], creating unphysical gaps at the bounds of the design variables range, thus hindering design updates.

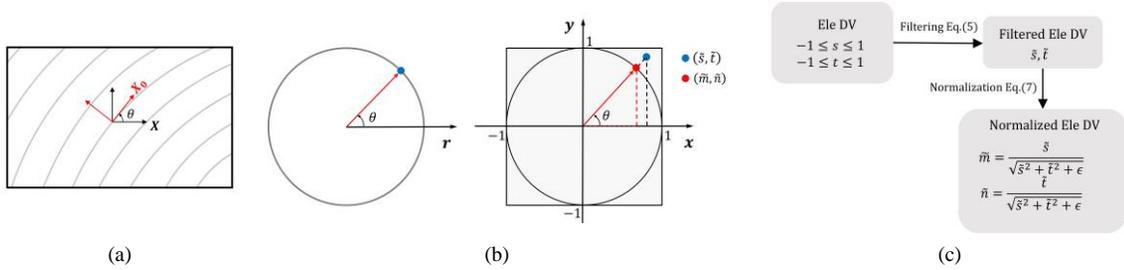

Fig. 1. (a) Illustration of the reference coordinate system ($X$, aligned with the discretization) and of the local one ($X_0$, aligned with the material orientation). (b) Schematic representation of the Polar (left), and Cartesian (right) representation of the same angle. (c) Illustration of the procedure going from the design variables $(s, t)$ to filtered design variables $(\tilde{s}, \tilde{t})$, and to the physical variables $(\tilde{m}, \tilde{n})$, used in the cartesian representation.

To circumvent these issues, in this work we adopt the so-called Cartesian representation [18], and we describe the orientation field by the unit vector field $\tilde{\boldsymbol{v}}(X_e) = \{\tilde{m}(X_e), \tilde{n}(X_e)\}^T$, where $\tilde{m}(X_e) = \cos(\theta(X_e))$ and $\tilde{n}(X_e) = \sin(\theta(X_e))$ (see Fig. 1(b), right). This is linked to the auxiliary field $\boldsymbol{\vartheta}(X_e) = \{s(X_e), t(X_e)\}^T$, where $s_e, t_e \in [-1,1]$ are the design variables that are optimized by the gradient-based algorithm.

The operations linking $\boldsymbol{\vartheta}(X_e)$ to $\tilde{\boldsymbol{v}}(X_e)$ are summarized in Fig. 1(c). First, the linear filtering [60]

$$\tilde{y}(X_e) = \frac{\sum_i w(X_i - X_e)y(X_i)}{\sum_i w(X_i - X_e)} \tag{4}$$

based on the distance-dependent kernel $w(X_i - X_e) = \max\{0, 1 - \|X_i - X_e\|r_f^{-1}\}$, where $r_f > 0$, is applied to both $s$ and $t$ separately, obtaining the filtered field $\tilde{\boldsymbol{\vartheta}}(X_e) = \{\tilde{s}(X_e), \tilde{t}(X_e)\}^T$. Then, this is projected to the physical orientation field with the normalization

$$\widetilde{\boldsymbol{v}} = \begin{bmatrix} \widetilde{m} \\ \widetilde{n} \end{bmatrix} = \begin{bmatrix} \dfrac{\tilde{s}}{\sqrt{\tilde{s}^2 + \tilde{t}^2 + \epsilon}} \\ \dfrac{\tilde{t}}{\sqrt{\tilde{s}^2 + \tilde{t}^2 + \epsilon}} \end{bmatrix} \tag{5}$$

where $\epsilon = 10^{-6}$ is used to avoid numerical singularities, possibly caused by vanishing orientation vectors.

Compared to the Polar representation, the Cartesian representation doubles the number of design variables used for representing the orientation field. However, this allows for an increased design freedom, mitigating the above-mentioned difficulties linked to the Polar representation [18,58].

By substituting Eq.(5) into Eq. (2) and (3), the transformation matrix and the material tensor become a function of the normalized orientation vector components $(\widetilde{m}, \widetilde{n})$

$$\boldsymbol{C}_x(\widetilde{m}, \widetilde{n}) = \boldsymbol{Q}_e(\widetilde{m}, \widetilde{n})^T \boldsymbol{C}_0 \boldsymbol{Q}_e(\widetilde{m}, \widetilde{n})$$

$$\boldsymbol{Q}_e(\widetilde{m}, \widetilde{n}) = \begin{bmatrix} \widetilde{m}^2 & \widetilde{n}^2 & \widetilde{m}\widetilde{n} \\ \widetilde{n}^2 & \widetilde{m}^2 & -\widetilde{m}\widetilde{n} \\ -2\widetilde{m}\widetilde{n} & 2\widetilde{m}\widetilde{n} & \widetilde{m}^2 - \widetilde{n}^2 \end{bmatrix} \tag{6}$$

## 3. Manufacturing constraints

A critical requirement for the tow steering process is that the fiber paths maintain turning radii that are larger than the smallest turning radius $R_{min}$ at which fiber tows can be placed without causing defects, such as local fiber wrinkling, or buckling [29,30]. This is equivalent to a maximum allowable condition on the curvature $\overline{\kappa}$ of the fiber paths, as turning radius is the inverse of curvature. A second fundamental manufacturing requirement involves the prevention of gaps and overlaps, which may occur when adjacent fiber tows deviate from their parallel alignment. These constraints must be accounted for during the numerical design optimization to prevent the optimizer from exploiting their absence and producing designs that may include discontinuities, gaps, overlaps and regions violating the minimum turning radius in the fiber paths. This in turn would require extensive post-processing operations, compromising the performance predicted by numerical optimization..

Following [36], tow curvature, and the tendency to form gaps/overlaps, can be related to the curl and divergence of the tow steered path, respectively. Recalling the vector field representation of the orientation path: $\widetilde{\boldsymbol{v}}(\boldsymbol{X}) = \widetilde{m}(\boldsymbol{X})\boldsymbol{i} + \widetilde{n}(\boldsymbol{X})\boldsymbol{j}$, its curl and divergence are defined as

$$\kappa(\boldsymbol{X}) := (\nabla_X \times \widetilde{\boldsymbol{v}}(\boldsymbol{X})) \cdot \boldsymbol{k} = \frac{\partial \widetilde{n}(\boldsymbol{X})}{\partial x} - \frac{\partial \widetilde{m}(\boldsymbol{X})}{\partial y} \tag{7}$$

$$\psi(\boldsymbol{X}) := \nabla_X \cdot \widetilde{\boldsymbol{v}}(\boldsymbol{X}) = \frac{\partial \widetilde{m}(\boldsymbol{X})}{\partial x} + \frac{\partial \widetilde{n}(\boldsymbol{X})}{\partial y} \tag{8}$$

Eq. (7) and (8) express the derivative of the angle field $\theta(\boldsymbol{X})$ along the tows' tangent and transversal directions, respectively. The first corresponds to the definition of the fibers' path curvature. The second can be used to identify the tendency to create gaps and overlaps due to the fiber paths. Indeed, if $\psi(\boldsymbol{X}) > 0$ the density of fibers entering the control volume at $\boldsymbol{X}$ is larger than that of fibers leaving the same volume. Thus, a gap is likely to form at some point. Likewise, if $\psi(\boldsymbol{X}) < 0$ the density of fibers entering the control volume is smaller than that of fibers leaving, and an overlap is likely to form.

Fig. 2(a) shows an example of an orientation field, with non-directional lines representing fiber orientations, and (b) and (c) show the contour plots corresponding to the magnitude of the fiber paths curvature ($|\kappa|$) and divergence ($|\psi|$), respectively. These magnitudes are evaluated from the discretized orientation field, using finite differences to approximate Eq. (7) and (8) (see Appendix B for details), and the fiber path patterns are visualized by generating near-evenly spaced streamlines. Fig. 2(b) clearly shows the correlation between large $|\kappa|$ and highly curved streamlines (*i.e.*, tows with small turning radius). Similarly, from Fig. 2(c), we see how large $|\psi|$ magnitudes correspond to regions where the fiber-path density increases and decreases, indicating deviations from parallel fiber paths, and potential formation of gaps or overlaps.

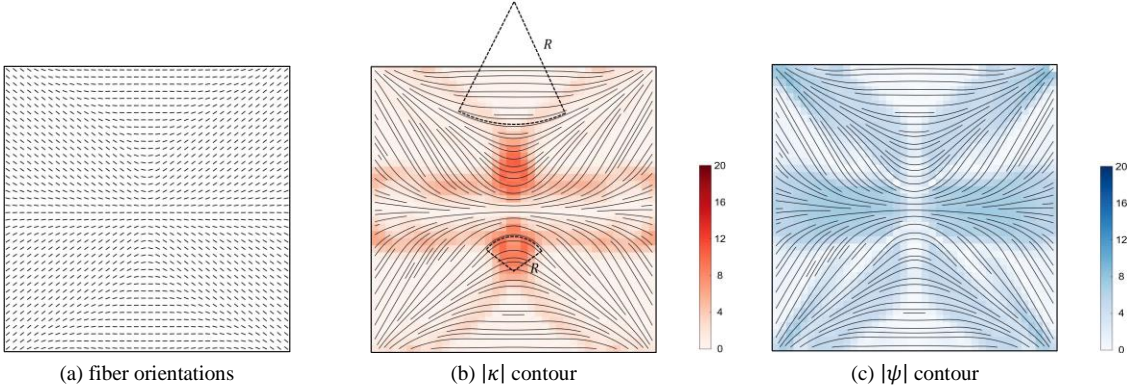

(a) fiber orientations     (b) $|\kappa|$ contour     (c) $|\psi|$ contour

Fig. 2. Illustration of a fiber orientation field on a unit square: (a) fiber orientations; and streamline plot with color contours showing the magnitude of the field's (b) curl and (c) divergence, as computed by Eq. (7) and (8), respectively. From these plots we clearly recognize the link between these operators and the turning radius ($1/|\kappa|$) and density of gaps/overlaps ($|\psi|$) in the tows' field.

The manufacturing constraints on minimum turning radius and maximum density of gaps/overlaps requirements can thus be enforced by locally bounding the magnitude of divergence and curl of the orientation field:

$$|\kappa(\boldsymbol{X})| \leq \overline{\kappa}$$
$$|\psi(\boldsymbol{X})| \leq \overline{\psi} \tag{9}$$

where $\overline{\kappa}$ and $\overline{\psi}$ are the maximum allowable curvature and divergence values, respectively. We highlight that the sign of both the curvature and divergence in Eq. (7) and (8) lack physical significance for the tow steering process, since $(\widetilde{m}, \widetilde{n})$ and $(-\widetilde{m}, -\widetilde{n})$, which yield $\kappa(\boldsymbol{X})$ and $\psi(\boldsymbol{X})$ with same magnitude and opposite sign, represent the same fiber orientation. Therefore, we impose uniform maximum bounds, denoted as $\overline{\kappa}$ and $\overline{\psi}$, for the unsigned curvature and divergence, respectively.

The values of $\overline{\kappa}$ and $\overline{\psi}$ are based on properties of the feedstock and manufacturing equipment. For example, carbon nanotube yarns tend to achieve very small turning radii [61]. The upper bound $\overline{\kappa}$ is usually determined experimentally by testing different radii of curvature with different combinations of process parameters such as temperature, roller speed, and roller pressure [62]. The upper bound $\overline{\psi}$ can be selected based on the minimum and maximum divergence values of a manufacturable fiber path derived in [36] as

$$\psi_{min} = -\frac{\ln\left(\frac{1+a_g}{2(1-a_o)}\right)}{L_{cut}}, \quad \psi_{max} = \frac{\ln\left(\frac{1+a_g}{2(1-a_o)}\right)}{L_{add}} \tag{10}$$

where $a_g$ and $a_o$ are predefined gaps and overlaps percentage limits, respectively. $L_{cut}$ is the minimum allowable tow cut length, which refers to the smallest length of tow that can be laid by the AFP machine before the tow can be cut. The variable $L_{add}$ is the minimum allowable tow addition length, which refers to the smallest length of tow that can be laid before the machine may consider adding an additional tow in the gap between two adjacent tow paths. For a more complete derivation of the relationship between divergence and density of gaps/overlaps percentage of a tow, the interested reader is referred to [36].

Finally, we point out that the magnitudes of curvature and divergence that can be represented over a given mesh depends on the mesh fineness. Equations (A.3) and (A.4), clearly imply $(|\kappa|, |\psi|) \leq 1/h_x + 1/h_y$ where $h_x$ and $h_y$ represent the element sizes in the $x$ and $y$ directions, respectively. Thus, achieving highly curved fiber paths, in the optimization process requires a rather fine discretization of the design domain. This aligns with similar considerations in work employing reference curves or level set functions to represent fiber paths, where fine mesh discretization is crucial for accurately identifying fiber orientations around sharp turns.

## 4. Optimization problem and solution strategy

We will demonstrate the maximum curvature and maximum density of gaps/overlaps constraints in the context of design for maximum structural stiffness, or the so-called minimum compliance problem. The optimization problem is expressed as:

$$\min_{(s,t)\in[-1,1]^{n_e}} c(s,t) = f^T u(s,t) \tag{11a}$$

$$\text{s.t.} \quad K(s,t)u(s,t) = f \tag{11b}$$

$$|\kappa^e(s,t)| - \overline{\kappa} \leq 0, \quad \forall\, e = 1, \dots n_e \tag{11c}$$

$$|\psi^e(s,t)| - \overline{\psi} \leq 0, \quad \forall\, e = 1, \dots n_e \tag{11d}$$

where $n_e$ is the number of finite elements, $f$ is the vector of applied nodal forces, and $u$ the vector of corresponding nodal displacements. Since we adopt a nested optimization solution strategy [63], the displacements are computed by explicitly solving the equilibrium equation (11b) at each re-design step, where $K(s,t)$ is the global stiffness matrix, depending on the design variables $(s,t) \in [-1,1]^{n_e}$.

### 4.1. Treatment of local constraints by the AL method

The optimization problem (11) is solved using the Method of Moving Asymptotes (MMA) [64,65], a primal-dual algorithm based on locally convex, separable approximations of the original objective and constraint functions. The manufacturing requirements of Eq. (11c-d) are applied to each finite element in the discretization, resulting in a large number $(n_e)$ of constraints. To treat these efficiently, without resorting to a constraint aggregation strategy [45], in this work we adopt the Augmented Lagrangian (AL) method.

The essential idea of the AL method is to replace the original optimization problem (11) with a sequence of subproblems, featuring the minimization of the AL function, with box constraints on the design variables. The original constraints are introduced in the AL function through a quadratic penalty term, which imposes a high cost to violated constraints. Specifically, at the $k$-th step of the AL method, the optimization subproblem is

$$\min_{(s,t)\in[-1,1]^{n_e}} L^{(k)}(s,t) = c(s,t) + \frac{\Lambda}{N} \sum_{j=1}^{N} \left( \lambda_j^{(k)} h_j^{(k)}(s,t) + \frac{\mu^{(k)}}{2} \left( h_j^{(k)}(s,t) \right)^2 \right) \tag{12}$$

$$\text{s.t.} \quad K(s,t)u(s,t) = f$$

where

$$h_j^{(k)} = \max\left( g_j, -\frac{\lambda_j^{(k)}}{\mu^{(k)}} \right) \tag{13}$$

$$g_j = \begin{cases} \dfrac{\kappa^e(\boldsymbol{s},\boldsymbol{t})}{\overline{\kappa}} - 1, & j = 1, \dots, n_e & (14a) \\ -\dfrac{\kappa^e(\boldsymbol{s},\boldsymbol{t})}{\overline{\kappa}} - 1, & j = n_e + 1, \dots, 2n_e & (14b) \\ \dfrac{\psi^e(\boldsymbol{s},\boldsymbol{t})}{\overline{\psi}} - 1, & j = 2 \times n_e + 1, \dots, 3n_e & (14c) \\ -\dfrac{\psi^e(\boldsymbol{s},\boldsymbol{t})}{\overline{\psi}} - 1, & j = 3 \times n_e + 1, \dots, 4n_e & (14d) \end{cases}$$

In the formulation of (12), $L^{(k)}(\boldsymbol{s},\boldsymbol{t})$ is the AL function at the *k*-th redesign step, and the second term penalizes constraint violation. $\Lambda$ is the parameter governing the contribution of constraints in the AL function. $\{\lambda_j^{(k)}\}_{j=1}^{N}$ is the vector of Lagrange multipliers, and $\mu^{(k)} > 0$ is the parameter governing the quadratic penalty. From the definition of $h_j^{(k)}$, in Eq. (13), the inactive constraints are not contributing to the AL function.

After solving the $k - th$ optimization subproblem, the Lagrange multiplier vector and the penalty parameter are updated as

$$\lambda_j^{(k+1)} = \lambda_j^{(k)} + \mu^{(k)} h_j^{(k)} \quad (15)$$

$$\mu^{(k+1)} = \min(\alpha \mu^{(k)}, \mu_{max}) \quad (16)$$

where $\alpha > 1$ such that $\mu^{(k)}$ will eventually exceed the threshold level for multiplier convergence [46]. The variable $\mu_{max}$ is the upper bound to prevent $\mu^{(k)}$ from being too large to avoid ill-conditioning issues.

The subproblem in equation (12) is solved by the MMA [65], scaling the objective function such that its initial value is between 1 and 100. The default parameter settings are summarized in Table 1. The initial Lagrange penalty parameter is set to 10, and is increased by a factor of 1.5 in each AL update until reaching the maximum value of $1e6$. In the authors' experience, the performance of the AL method is very sensitive to the value of $\Lambda$. If $\Lambda$ is too large, the AL function is dominated by the penalty term, overly restricting the optimization progresses. If $\Lambda$ is too small, the quadratic penalty term has little influence, potentially leading constraint violations. Therefore, we initiate $\Lambda$ to a relatively small value ($\Lambda^{(0)} = 0.01$), allowing for some violation of the manufacturing constraints in the early optimization stages; then we increase it by an updating factor of 1.3 in each AL update until reaching the maximum value of $\Lambda_{max} = 1$, ensuring fulfilling (or very minor violations) of the manufacturing constraints in the final designs.

Table 1. Default parameter magnitudes for the parameters used in the AL function, and for the MMA optimizer, adopted for all the following numerical examples.

| Parameter | Value |
| --- | --- |
| regularization parameter, $\Lambda^{(0)}$ | 1e-2 |
| maximum regularization parameter, $\Lambda_{max}$ | 1 |
| updating factor for regularization parameter | 1.3 |
| initial penalty parameter, $\mu^{(0)}$ | 1e1 |
| maximum penalty parameter, $\mu_{max}$ | 1e6 |
| updating factor for penalty parameter, $\alpha$ | 1.5 |
| initial Lagrange multiplier, $\lambda_j^{(0)}$ | 0 |
| MMA iterations per AL step | 20 |
| maximum number of iterations | 1000 |
| move limit per MMA iteration | 0.05 |

The sensitivity analysis for the objective function and the constraints in (12), and thus also for the terms in the AL (14), is given in Appendix B.

## 5. Numerical examples

We demonstrate the proposed method to design structures with maximized stiffness. Three benchmark problems are considered: the L-bracket structure, a simply supported beam, and a cylindrical shell structure (Fig. 3). We assume base material properties of a carbon fiber epoxy composite with elastic moduli $E_1 = 140$ GPa, $E_2 = E_3 = 9.5$ GPa, $G_{12} = G_{13} = G_{23} = 5.8$ GPa, and $\nu_{12} = \nu_{13} = \nu_{23} = 0.3$ for the L-bracket and cylindrical shell examples, adopting the constitutive laws of Section 2 and Appendix A, respectively. For the simply supported beam problem we consider $E_1 = 100$ GPa, $E_2 = E_3 = 5$ GPa, $G_{12} = G_{13} = G_{23} = 3$ GPa, and $\nu_{12} = \nu_{13} = \nu_{23} = 0.3$. The L-bracket and the simply supported beam are modelled with plane stress assumptions and discretized by four-node quadrilateral elements. The cylindrical shell problem is discretized by a 3D shell element, specifically developed for laminated composite shell structures (see details in [66,67]).

Following [36], depending on feedstock properties and manufacturing equipment, the minimum turning radius $R_{min}$ can typically vary between [0.02,1]m and the minimum cut/add length ($L_{cut} = L_{add}$) between [0.04,4] m. Therefore, Equation (10) gives the corresponding divergence and curvature limits in the range $\bar{\kappa} \in$ [1,50] (1/m), and $\bar{\psi} \in [0.1,10]$ (1/m).

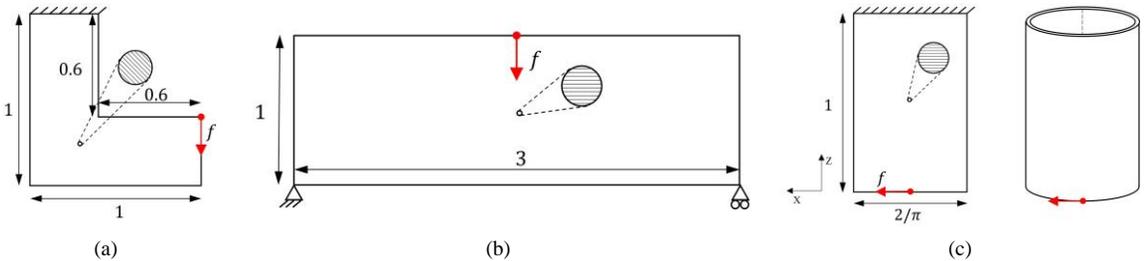

(a)          (b)          (c)

Fig. 3. Geometry and boundary conditions for the three benchmark examples: (a) L-bracket, (b) simply supported beam, and (c) cylindrical shell (thickness = 0.01 m). For each example, the initial fiber orientations used in the optimization are shown in the circular markers. Force and pinned supports, while depicted pointwise, are in practice applied over a small area spanning three finite elements, to prevent stress concentrations.

## 5.1. L-bracket

With this example, we investigate the impact of the manufacturing constraints on the optimized design and its performance, and we compare the proposed approach with that based on orientations filtering. To this end, we begin with a traditional continuous fiber angle optimization, only applying the linear filtering of Eq. (4) to the orientation design variables, and no manufacturing constraints. Next, we separately investigate the impact of maximum density of gaps/overlaps and maximum curvature constraints, while considering their indirect influence on each other. Finally, we achieve manufacturable designs by applying both constraints.

We discretize the design domain in Fig. 3 (a) by a 40×40 mesh. The initial design features unidirectional fiber at $\theta = -45°$ (i.e., $s_e = 0.5$ and $t_e = -0.5$) corresponding to a compliance energy of $C(s,t) = 144.64$ J.

### Continuous fiber angle optimization (CFAO) without manufacturing constraints

Fig. 4 collects minimum compliance L-bracket designs for increasing orientation filter radii. Fibers are generally aligned with the maximum principal stress direction; however, for small filtering sizes (e.g., $r_f = 0.05$m in Fig. 4(a)), we see rapid variations between fiber orientations, including with variances of up to 90°. Even if these variations and discontinuities are functional to compliance minimization, they create stress concentrations, and therefore should be avoided.

As the orientation filtering size increases, fiber paths become smoother. However, some rapid variations persist, especially in regions highlighted by the red circles. Fiber continuity is achieved by a large filtering size of $r_f = 2.0$m (e), at the cost of a large compliance increase. This filter size relative to the domain size is significantly larger than used in typical topology optimization problems. We indeed notice some clear signs of non-optimality; for instance, the fiber paths in the upper part of L-bracket, which in Fig. 4(d) are at 90° (as expected), become more tilted in Fig. 4(e) due to the larger filter size.

Table 2 summarizes the performance of designs obtained without and with manufacturing constraints. Under no manufacturing constraints, increasing the filtering size consistently reduces the maximum curvature but does not necessarily decrease the maximum divergence. For instance, in the divergence plot of Fig. 4(d), the blue-colored region in the bottom right corner exhibits a high divergence value of $max(|\psi|) = 74$ (1/m), indicating a significant tendency for fiber paths to cluster or disperse, which can complicate manufacturing and introduce defects. With a very large filtering size in Fig. 4(e), maximum curvature and maximum divergence decrease to 2.6 (1/m) and 2.7 (1/m), respectively. Nevertheless, achieving precise control and customization of these magnitudes for varying manufacturing requirements remains a challenge when using filtering.

| Fiber orientation | Principal stress directions | $|\kappa|$ | $|\psi|$ |
|---|---|---|---|
| 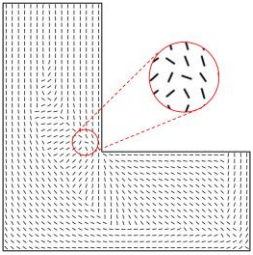 | 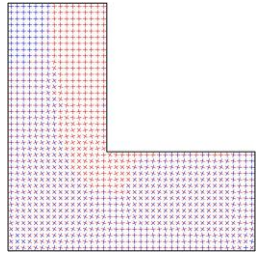 | 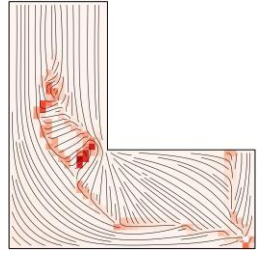 | 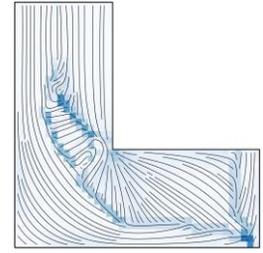 |

(a) $r_f = 0.05$m

| | | | |
|---|---|---|---|
| 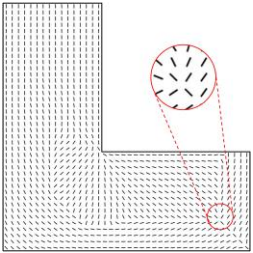 | 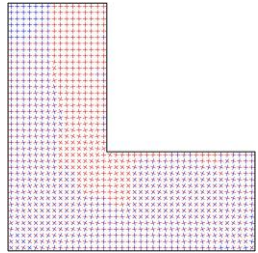 | 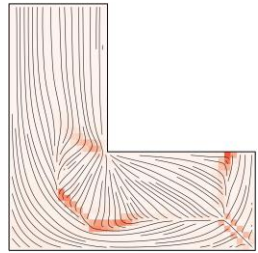 | 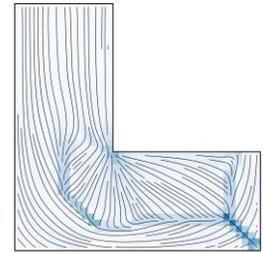 |

(b) $r_f = 0.125$m

| | | | |
|---|---|---|---|
| 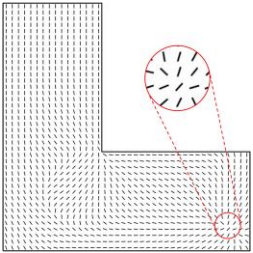 | 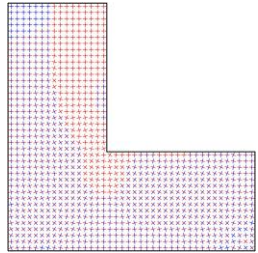 | 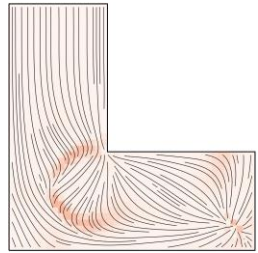 | 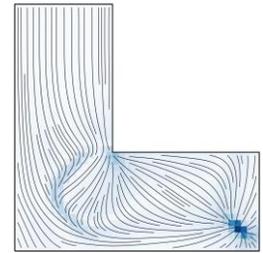 |

(c) $r_f = 0.25$m

| | | | |
|---|---|---|---|
| 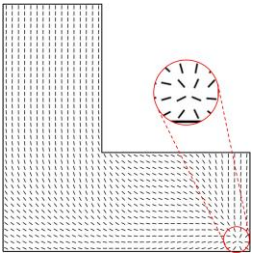 | 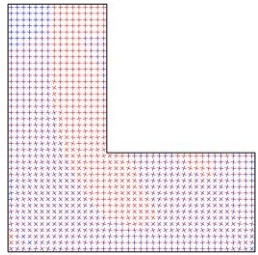 | 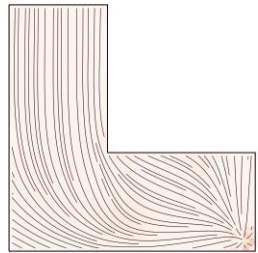 | 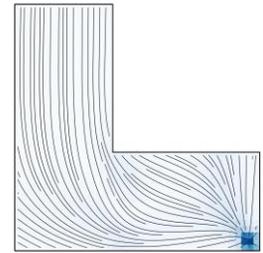 |

(d) $r_f = 0.375$m

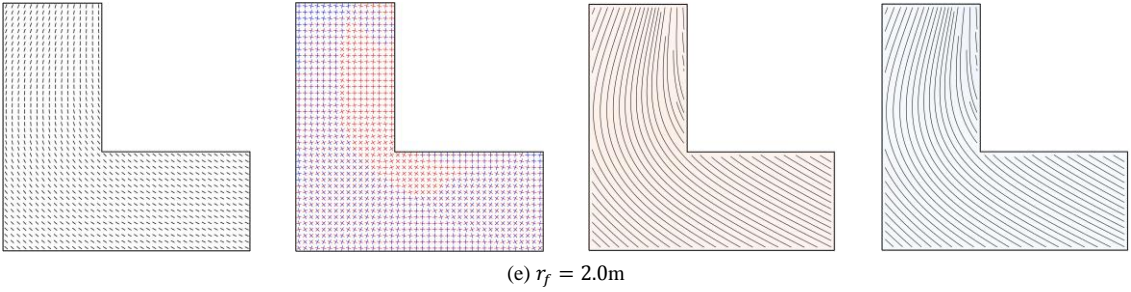

(e) $r_f = 2.0$m

Fig. 4. Optimized L-bracket designs obtained without considering manufacturing constraints and for increasing value of the orientation filtering radius ($r_f$). From left to right: fiber orientations, principal stresses directions, streamlines plot with field distribution of the curvature $|\kappa|$, and divergence field $|\psi|$. The principal stresses are evaluated at the center of each element: red lines indicate tensile stress and blue lines indicate compressive stress. Compliance magnitudes have units of J, and curvature and divergence of (1/m), which will be used in all of the figures that follow.

*Gaps/overlaps or curvature control only*

We now incorporate manufacturing requirements concerning fiber path gaps/overlaps and curvature into the design optimization formulation, while keeping a minimum orientation filtering radius $r_f = 0.05$m. We first apply only the divergence constraint with progressively stricter bounds: $\overline{\psi} = [10, 5, 2, 1, 0.5, 0.2, 0.1]$ (1/m). Fig. 5 presents the resulting optimized designs, including fiber path patterns and divergence contour plots. When a small tow cut/add length is allowed, the optimized design still exhibits large variation in fiber orientations (see Fig. 5 (a)). Moreover, the streamline plot in this case reveals multiple tow cuts/adds within the design domain, indicating gap/overlap formation. As the minimum allowed cut/add length increases, i.e., the divergence constraints become more restrictive, fiber path patterns exhibit fewer cuts/adds, aligning fibers more closely with their neighbors (see Fig. 5 (b-g)). Moreover, we notice how the fiber orientations are regularized and smoothed without significant changes between neighboring elements, ensuring fiber continuity. Thus, a stricter divergence constraint has an indirect effect also on the curvature radius of the tows.

When the divergence constraint limit is decreased to $\overline{\psi} = 0.1$ (1/m) the optimized fiber path pattern achieves the highest quality manufacturability in terms of the amount of gaps/overlaps for tow steering (see Fig. 5(g)). However, this simplified design comes at the cost of an approximately 8.2% increase in compliance compared to the design at $\overline{\psi} = 1.0$ (1/m) in Fig. 5(d). It is noticed that, compared to the solution obtained with even a larger filter in Fig. 4(e), the fiber paths at the upper region in Fig. 5(g) are kept basically straight. This explains the better performance of this design, indicated by a lower compliance shown in Table 2. The maximum violation of divergence constraints in all designs is less than 5%, see Table 2.

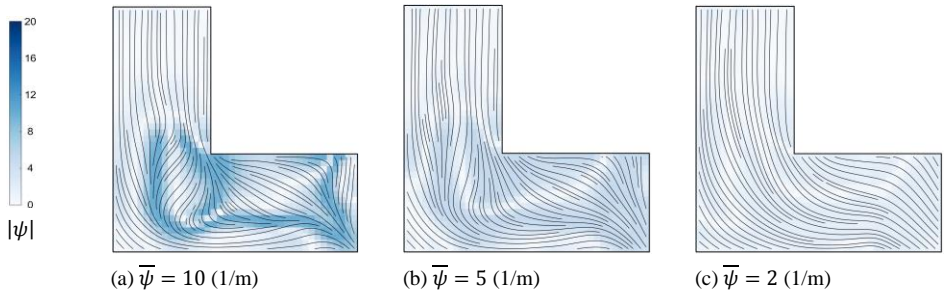

(a) $\overline{\psi} = 10$ (1/m)    (b) $\overline{\psi} = 5$ (1/m)    (c) $\overline{\psi} = 2$ (1/m)

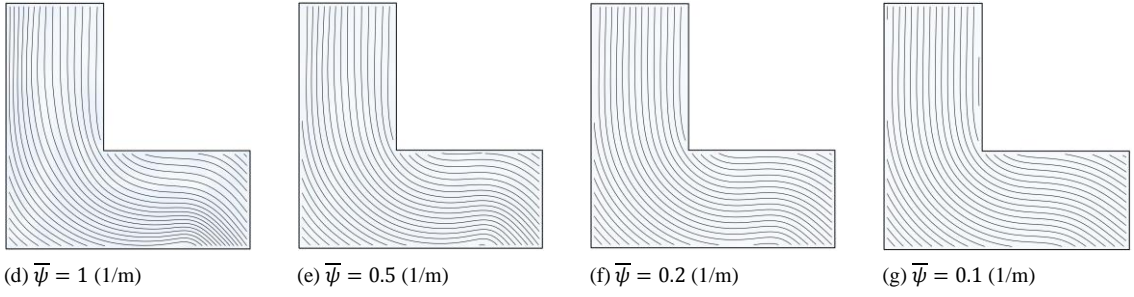

(d) $\overline{\psi} = 1$ (1/m)    (e) $\overline{\psi} = 0.5$ (1/m)    (f) $\overline{\psi} = 0.2$ (1/m)    (g) $\overline{\psi} = 0.1$ (1/m)

Fig. 5. Optimized L-bracket designs obtained considering divergence constraints to control the density of gaps/overlaps. The colormaps, showing the distribution of the maximum divergence of the orientation field, are all referred to the range shown in the top left corner. The divergence constraint is active in all the cases, and slightly violated (less than 5%) for case (f).

Table 2. Results for L-bracket designs without manufacturing constraints(left), with only divergence constraints (middle), and with only curvature constraints (right). Compliance magnitudes have units of J, and curvature and divergence of (1/m), which will be used in all of the tables that follow.

| No manufacturing constraints | | | | Gaps/Overlaps Control Only | | | | Curvature Control Only | | | |
|---|---|---|---|---|---|---|---|---|---|---|---|
| $r_f$ | $max(|\kappa|)$ | $max(|\psi|)$ | $C$ | $\overline{\psi}$ | $max(|\kappa|)$ | $max(|\psi|)$ | $C$ | $\overline{\kappa}$ | $max(|\kappa|)$ | $max(|\psi|)$ | $C$ |
| 0 | 51.5 | 58.4 | 28.24 | 10 | 26.4 | 10.0 | 28.99 | 50 | 40.5 | 58.2 | 25.38 |
| 0.05 | 66.2 | 59.5 | 26.69 | 5 | 13.3 | 5.0 | 34.73 | 20 | 20.0 | 68.2 | 25.70 |
| 0.125 | 44.2 | 58.5 | 25.58 | 2 | 7.1 | 2.0 | 40.59 | 10 | 10.0 | 35.7 | 27.06 |
| 0.25 | 21.7 | 74.7 | 27.58 | 1 | 12.4 | 1.0 | 43.21 | 5 | 5.0 | 33.4 | 30.09 |
| 0.375 | 15.8 | 74.0 | 35.61 | 0.5 | 14.9 | 0.50 | 44.72 | 2.5 | 2.50 | 22.6 | 34.05 |
| 2.0 | 2.6 | 2.7 | 56.91 | 0.2 | 9.1 | 0.21 | 45.88 | 1.67 | 1.70 | 98.2 | 38.56 |
| | | | | 0.1 | 6.0 | 0.10 | 46.74 | 1 | 1.00 | 15.8 | 47.55 |

We now examine the impact of curvature control by enforcing only curvature constraints with progressively tighter limits: $\overline{\kappa} = [50, 20, 10, 5, 2.5, 1.67, 1]$ (1/m), corresponding to gradually stricter allowable tow steering radii. Fig. 6 presents the optimized designs, including fiber path patterns and curvature contour plots.

As the curvature constraint becomes more restrictive from Fig. 6 (a) to (g), fiber orientations are regularized and smoothed, reducing variance between neighboring elements. Additionally, in Fig. 6 (a), the region near the bottom edge and loading point exhibits significant curvatures. With stricter curvature constraints the fiber path exhibits smaller curvatures throughout the design domain, as clearly shown by the curvature contour. For all designs shown in Fig. 6, the minimum turning radius in the orientation field complies with the imposed radius (see dashed sector-shaped lines); we have only a slight violation of the constraint for the design in Fig. 6 (f).

We note that restricting the curvature of the fiber paths does not necessarily reduce the maximum density of gaps/overlaps in the optimized design. As shown by Fig. 6 (g), even with tight curvature constraints, the fiber path pattern still exhibits numerous short tow cuts/adds, indicating the formation of gaps/overlaps. Therefore, the indirect effect of one manufacturing constraint on the other, previously observed when imposing maximum divergence control, does not hold for the case of maximum curvature control.

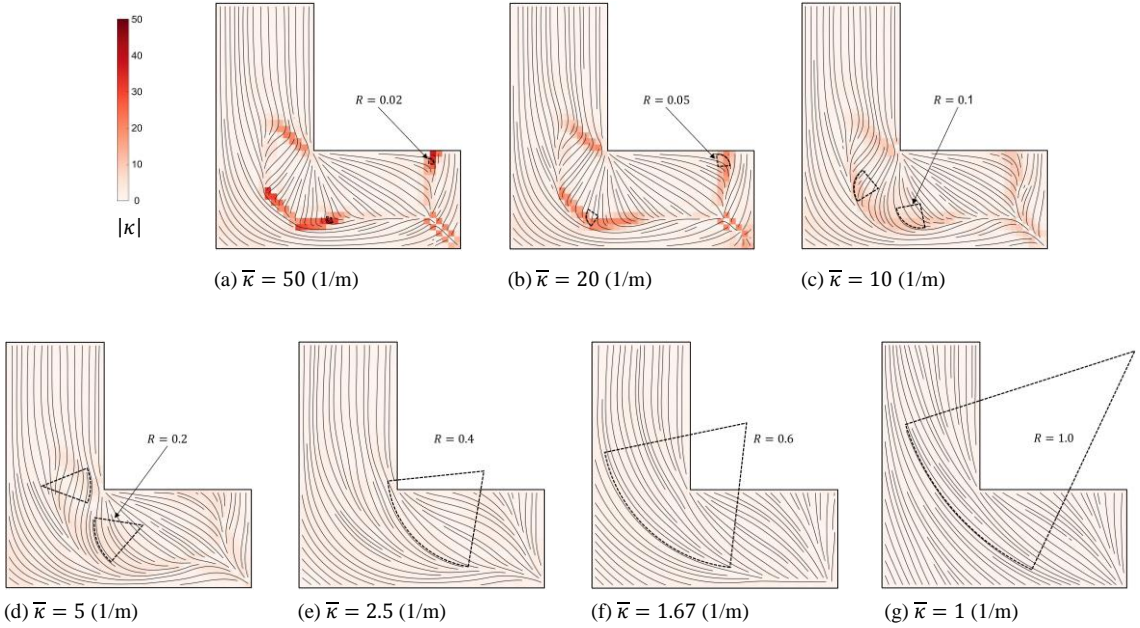

Fig. 6. Optimized L-bracket designs obtained considering curvature constraints to control the minimum steering radius of the tows (see the sector-shaped dashed line). The colormaps, showing the distribution of the curvature of the orientation field, all refer to the range shown in the top left corner. The constraint on maximum curvature is active in all the cases, and slightly violated (less than 2%) for case (f).

*Simultaneous gaps/overlaps and curvature control*

As shown in Table 2, we can achieve designs with improved manufacturability in terms of either tow gaps/overlaps or curvature. However, if the manufacturing facility has a strict requirement for both minimum turning radius and minimum cut/add lengths, e.g. $\overline{\kappa} = 2.5$ (1/m) and $\overline{\psi} = 2.5$ (1/m), these would still be difficult to manufacture. Therefore, it is crucial to incorporate both gap/overlap and curvature control.

For this study, we consider two sets of manufacturing constraints. The first set is derived from the design in Fig. 4(e), which exhibits the best design manufacturability among designs achieved without manufacturing constraints and with only filtering schemes. We adopt its maximum magnitudes of divergence and curvature and set constraint limits of $\overline{\kappa} = 2.5$ (1/m), $\overline{\psi} = 2.5$ (1/m). The second set uses the same minimum tow turning radius as the first set but imposes much stricter minimum tow cut/add length constraints. The corresponding applied constraint limits are $\overline{\kappa} = 2.5$ (1/m), $\overline{\psi} = 0.25$ (1/m).

Fig. 7 displays the two optimized designs with simultaneous control over curvature and gaps/overlaps. Both designs satisfy the minimum allowable turning radius of the fiber path, indicated by the dashed sector-shaped lines. In both designs, curvature and divergence constraints reach the maximum allowable limits, with a maximum violation of less than 1%.

Compared to the design in Fig. 4(e), the design in Fig. 7(a) shows similar manufacturability features, but significantly improved structural performance, with an almost 30% lower compliance. This demonstrates that the proposed method outperforms filtering schemes by providing precise control over tow manufacturing constraints while sacrificing less design freedom in optimizing fiber orientation for tow-steered composites.

By applying more restrictive divergence constraints than Fig. 7(a), the design in Fig. 7(b) features a reduced number of tow cut/add segments. Fiber paths align better with their neighbors and are more evenly spaced, resulting in a reduction in gaps/overlaps. As expected, this improved manufacturability in tow gaps/overlaps comes at the cost of increased structural compliance.

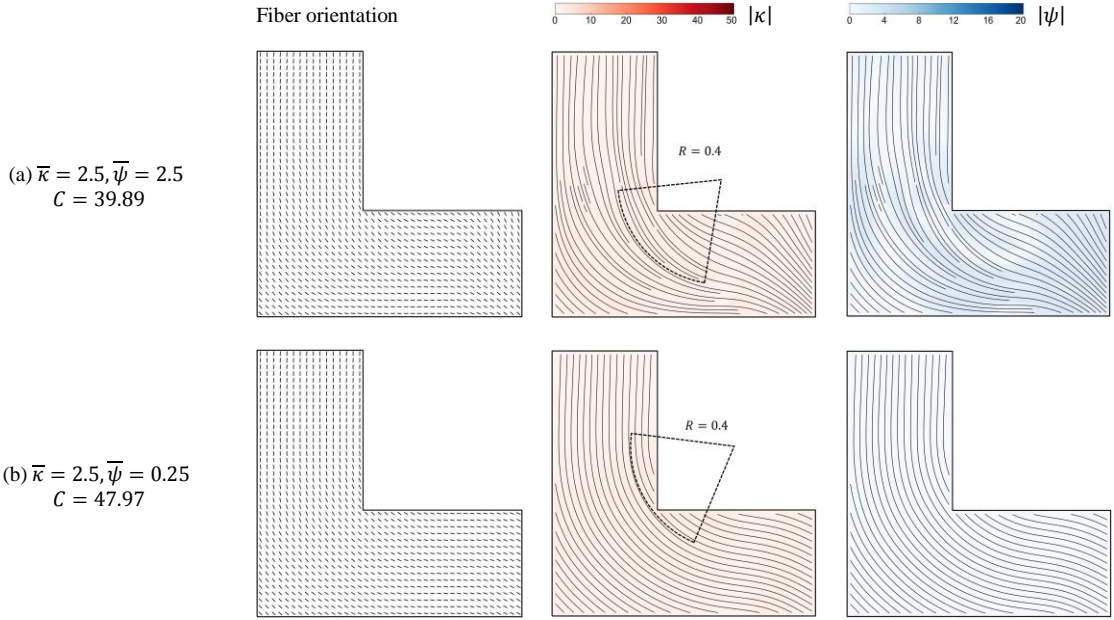

Fig. 7. Optimized designs obtained considering both curvature constraints and divergence constraints: fiber orientation distribution (left), streamlines plot with curvature field $|\kappa|$ (middle), and streamlines plot with divergence field $|\psi|$ (right).

The convergence history plots for both designs in Fig. 7 are depicted in Fig. 8(a) and (b). These plots illustrate a consistent decrease in structural compliance during the initial optimization stages, while allowing some violation of manufacturing constraints. This strategy allows the optimizer to explore the infeasible design space to minimize structural compliance as much as possible. Then, as the Augmented Lagrangian parameters are updated at each continuation step, the constraint violations decrease, at a (small) expense of structural performance. Eventually, the constraints become fulfilled after approximately 400 iterations.

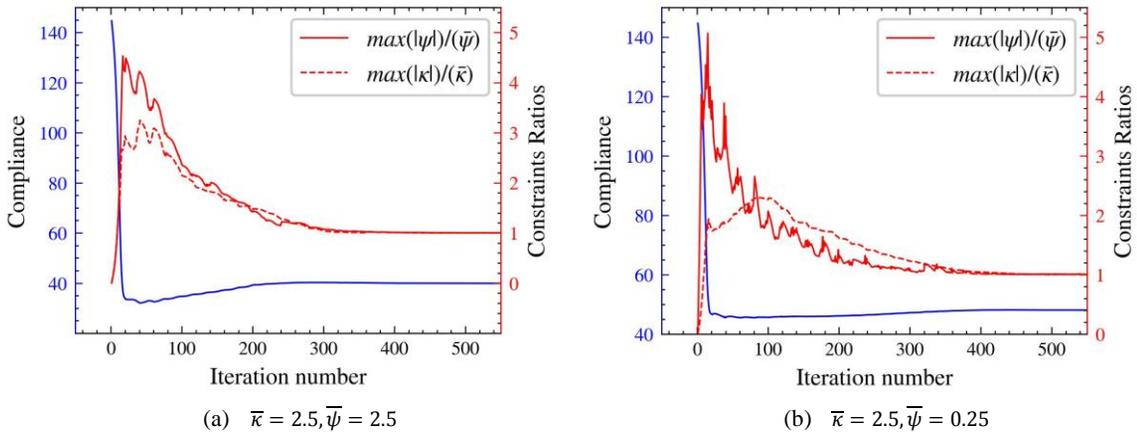

(a) $\bar{\kappa} = 2.5, \bar{\psi} = 2.5$  (b) $\bar{\kappa} = 2.5, \bar{\psi} = 0.25$

Fig. 8. Convergence plots for the designs in Fig. 7. The evolution of the structural compliance is plotted against the left axis, and that of manufacturing constraints ratios is plotted against the right axis. Compared to the initial design with unidirectional fibers, the designs in Fig. 7(a) and Fig. 7(b) reduce compliance (increase stiffness) by about 72% and 67%, respectively.

## 5.2. Simply supported beam

We discretize the design domain in Fig. 3(b) by a 90×30 mesh. The initial design features unidirectional fiber orientations at $\theta = 0°$ (i.e., $s_e = 1.0$ and $t_e = 0.0$) corresponding to $C(s,t) = 156.53$ J. A small orientation filtering radius $r_f = 1/6$m is applied for all cases with manufacturing constraints applied.

The optimized designs obtained for increasing values of the orientation filter radius, and without manufacturing constraints, are shown in Fig. 9. For these solutions, fibers near the bottom edge of the beam mainly align at $\theta = 0°$ to provide maximum resistance to tensile stresses, whereas in the upper part of the domain fibers align with the compression path connecting the loaded and support regions. When no orientation filtering or filtering with a small radius is applied, a sharp transition between these two regions occurs, leading to extensive regions with gaps/overlaps and high curvature of the tows (see Fig. 9(a) and (b)).

The maximum value of $|\psi|$ (but not that of $|\kappa|$) is reduced only for the quite large value of the orientations filter radius $r_f = 2/3$m. Then, for $r_f = 2.0$m we achieve a smooth orientation field (see Fig. 9(d)) with both maximum curvature and maximum divergence dropping to 2.7 (1/m) and 2.9 (1/m), and with a compliance increase of about 264% with respect to the compliance achieved for $r_f = 0$m (see Table 3).

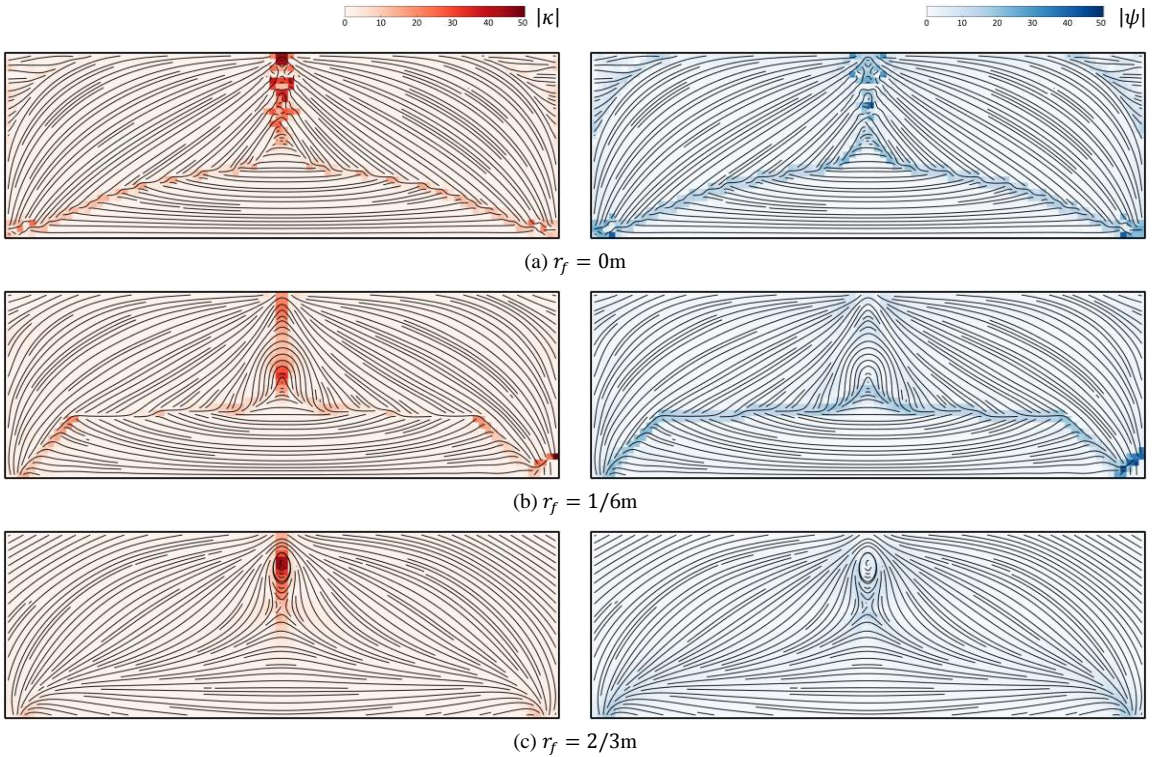

(a) $r_f = 0$m

(b) $r_f = 1/6$m

(c) $r_f = 2/3$m

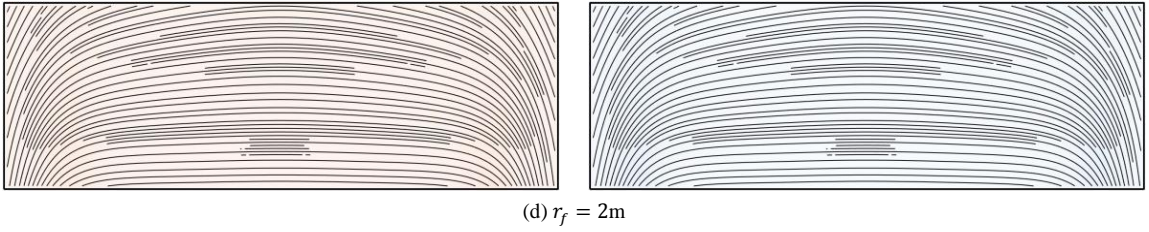

(d) $r_f = 2m$

Fig. 9. Optimized designs for the simply supported beam, obtained without enforcing manufacturing constraints and with increasing values of the orientation filter radius.

The performance of the designs obtained by imposing manufacturing constraints through the AL method, considering some combinations of $\bar{\kappa}$ and $\bar{\psi}$, are presented in Table 3, and two representative designs are shown in Fig. 10. In all these cases, both the curvature and gaps/overlaps constraints are active, and the maximum constraint violation for the final design is less than 0.5%.

Already the design corresponding to $\bar{\kappa} = 20$ (1/m) and $\bar{\psi} = 10$ (1/m) shows that for the same value of compliance, a design with much better manufacturability is achieved as compared to orientation filtering with $r_f = 2/3m$ and no manufacturing constraints, since both $\bar{\kappa}$ and $\bar{\psi}$ are more than halved. Moreover, none of the designs obtained by using orientation filtering alone can achieve magnitudes of $\bar{\kappa} \leq 2.5$ (1/m) or $\bar{\psi} \leq 2.5$ (1/m), as obtained by direct imposition of the manufacturing constraint (see Table 3). This comes at the cost of a much larger optimized compliance. However, if the beam design is not practically manufacturable, its numerically predicted performance may not be achievable in practice. We also note the compliance of the constrained design is less than the compliance of the design found when applying only orientation filtering.

Table 3. Results for the simply supported beam optimization without manufacturing constraint and with manufacturing constraints handled either by the AL method proposed in this study, or with the KS aggregation method.

| No manufacturing constraints | | | | Manufacturing constraints (AL method / KS aggregation) | | | | |
| --- | --- | --- | --- | --- | --- | --- | --- | --- |
| $r_f$ | $max(|\kappa|)$ | $max(|\psi|)$ | $C$ | $\bar{\kappa}$ | $\bar{\psi}$ | $max(|\kappa|)$ | $max(|\psi|)$ | $C$ |
| 0 | 44.5 | 41.5 | 37.37 | 20 | 10 | 20.0 / 18.8 | 10.0 / 11.7 | 42.93 / 48.42 |
| 1/6 | 56.8 | 43.1 | 36.25 | 2.5 | 2.5 | 2.50 / 2.92 | 2.50 / 2.67 | 78.89 / 78.52 |
| 2/3 | 46.2 | 17.1 | 43.76 | 1.0 | 2.5 | 1.00 / 1.14 | 2.51 / 2.85 | 82.64 / 92.89 |
| 2.0 | 2.7 | 2.9 | 98.78 | 2.5 | 0.25 | 2.50 / 2.88 | 0.25 / 0.28 | 96.47 / 119.1 |

Taking the combination $\bar{\kappa} = \bar{\psi} = 2.5$ (1/m) as a reference, we then consider two cases, tightening only one of the two constraints. When tightening the maximum curvature constraint to $\bar{\kappa} = 1.0$ (1/m), the optimized fiber path layout, shown in Fig. 10(a), resembles an arch shape, efficiently transferring loads to the supports. Compared to the design corresponding to $\bar{\kappa} = \bar{\psi} = 2.5$ (1/m), the curvature reduction is achieved with a relatively low increase of compliance (4%). When imposing a much stricter minimum tow cut/add length (thus, $\bar{\psi} = 0.25$ (1/m)), the optimized design, shown in Fig. 10(b), is completely free of gaps/overlaps. This comes at a larger increase in the compliance (24%), compared to the design with $\bar{\kappa} = \bar{\psi} = 2.5$ (1/m).

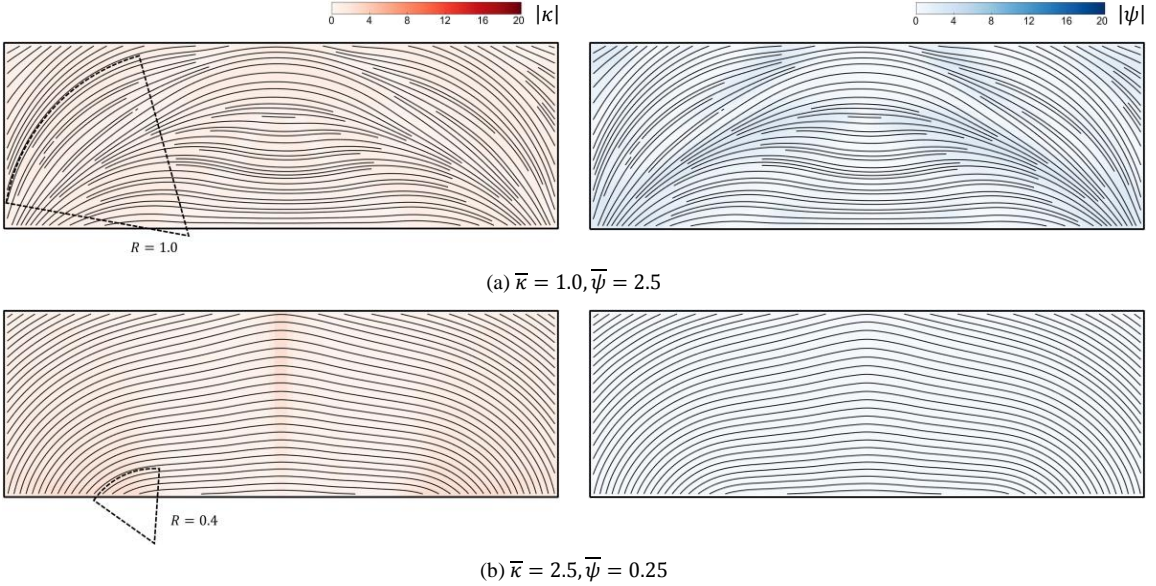

(a) $\bar{\kappa} = 1.0, \bar{\psi} = 2.5$

(b) $\bar{\kappa} = 2.5, \bar{\psi} = 0.25$

Fig. 10. Optimized designs for the simply supported beam, considering two sets of maximum curvature and gaps/overlaps constraints ($\bar{\kappa}, \bar{\psi}$). The AL method described in Section 4 is used to handle the many local constraints.

Finally, we compare the proposed method with the use of an aggregation strategy for the local manufacturing constraints. We use the Kreisselmeier–Steinhauser (KS) function [40] to replace Eq. (14) with a single global constraint

$$g_{KS}(p, g_j) = \frac{1}{p} \ln\left(\frac{1}{4n_e} \sum_{j=1}^{4n_e} \exp(p g_j)\right) \qquad (17)$$

where a continuation scheme is used to increase the aggregation parameter $p$, starting from $p = 5$ and scaling it by a factor of 1.1 at every 20 iterations, until reaching $p_{max} = 50$.

Designs obtained by constraint aggregation show the same or larger structural compliance than those achieved by the AL method (see Table 3). In particular, the designs obtained with the KS aggregation, for $\bar{\kappa} = 2.5$ (1/m) and $\bar{\psi} = 0.25$ (1/m) (see Fig. 11), show a 12.4% and 23.5% larger compliance compared to those in Fig. 11. This is a consequence of the inherent restriction of design flexibility for constraint aggregation methods, combined with the MMA optimizer, due to a much stronger penalization of the constraint violation in the early optimization stages (see Fig. 12). The peak constraints violation when using the KS aggregation method is approximately 100%, whereas we reach almost 470% constraint violation in the early stages of the optimization solved by the AL method, granting more design freedom. Also, the aggregation method results in a slight degree of constraint violation in the final design (see Table 3). These violations are likely mitigated by tunning parameters or using other aggregation functions, which is beyond the scope of the current work.

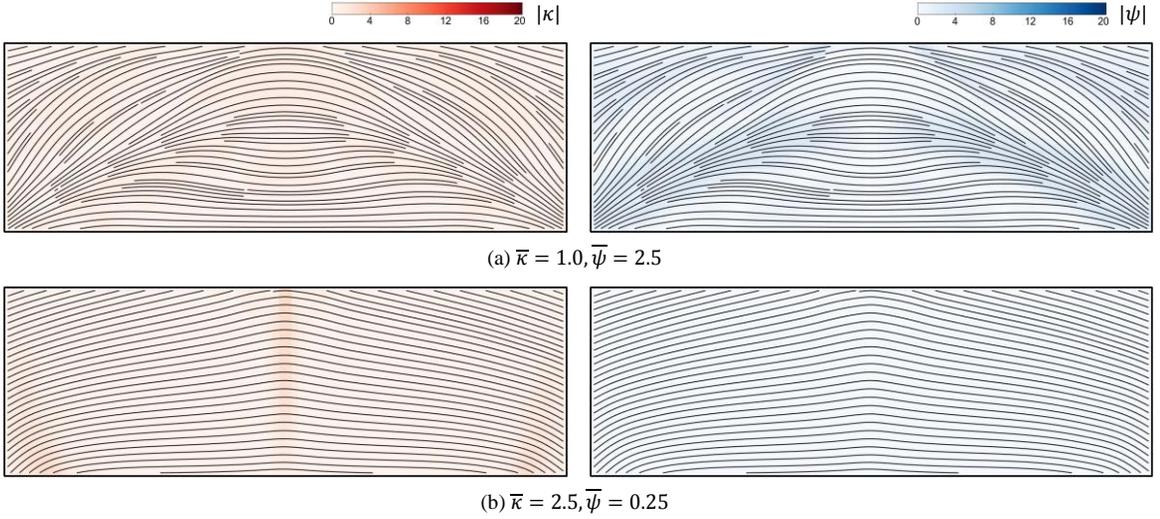

(a) $\overline{\kappa} = 1.0, \overline{\psi} = 2.5$

(b) $\overline{\kappa} = 2.5, \overline{\psi} = 0.25$

Fig. 11. Optimized designs for the simply supported beam, obtained considering two sets of maximum curvature and gaps/overlaps constraints ($\overline{\kappa}, \overline{\psi}$). The KS aggregation of Eq. (17) is used to handle the many local constraints.

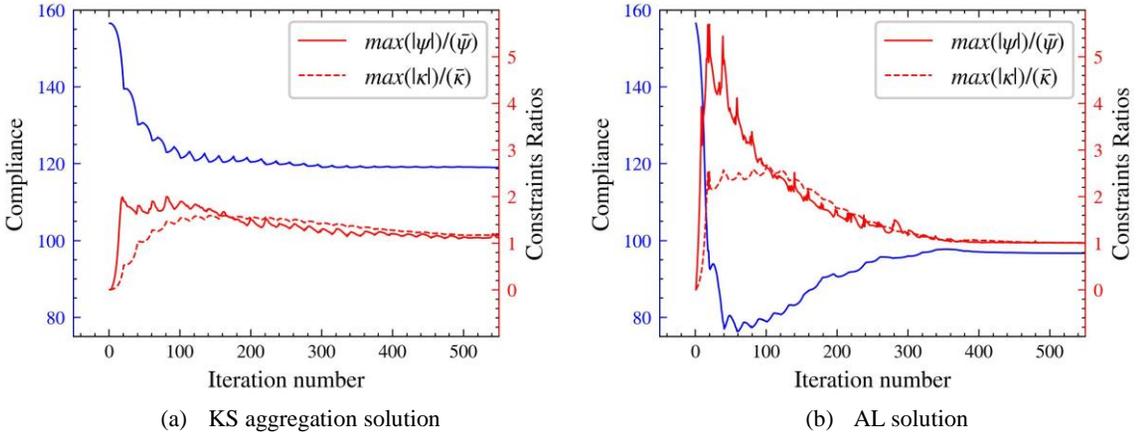

(a)  KS aggregation solution

(b)  AL solution

Fig. 12. Convergence plots for the designs of Fig. 11 (b) and Fig. 10 (b). Structural compliance is plotted against the left axis, and manufacturing constraints ratios are plotted against the right axis.

## 5.3. Cylindrical shell

For the cylindrical shell example of Fig. 3(c), the design variables are defined on a 2D rectangular plane and are then mapped on the cylindrical surface. The rectangular plane is obtained by cutting the cylindrical surface along the long-dashed line in Fig. 3(c) and then flattening it, while ensuring that the design variables on both sides of the cutting line are jointly considered during the evaluation of manufacturing constraints. The design domain is discretized into a grid of 40×20 elements. In the following we first analyze the design of a single ply, then extend to multi-ply design.

*Single-ply design*

The initial design features unidirectional fiber orientations $\theta = 0°$ (*i.e.*, $s_e = 1.0$ and $t_e = 0.0$) corresponding to $C(s,t) = 31.87$J. Fig. 13 shows the optimized fiber paths for two different filtering sizes: $r_f = 0.25$m and $r_f = 1$m, when manufacturing constraints are not introduced. The design is nearly symmetric about the vertical middle line, where loading is applied. The fiber path pattern gradually inclines upwards from the loading point towards both sides of the middle line, to efficiently distribute the load to the clamped boundary. It then turns and inclines gradually downwards until reaching the cutting lines.

Examining the designs in Fig. 13(a) and (b), we see that, even for the large filtering radius $r_f = 1$m, tow continuity is violated, particularly for the tows near the middle and cutting lines, exhibiting high curvature and sharp turns. This poses a significant manufacturing challenge for tow steering processes.

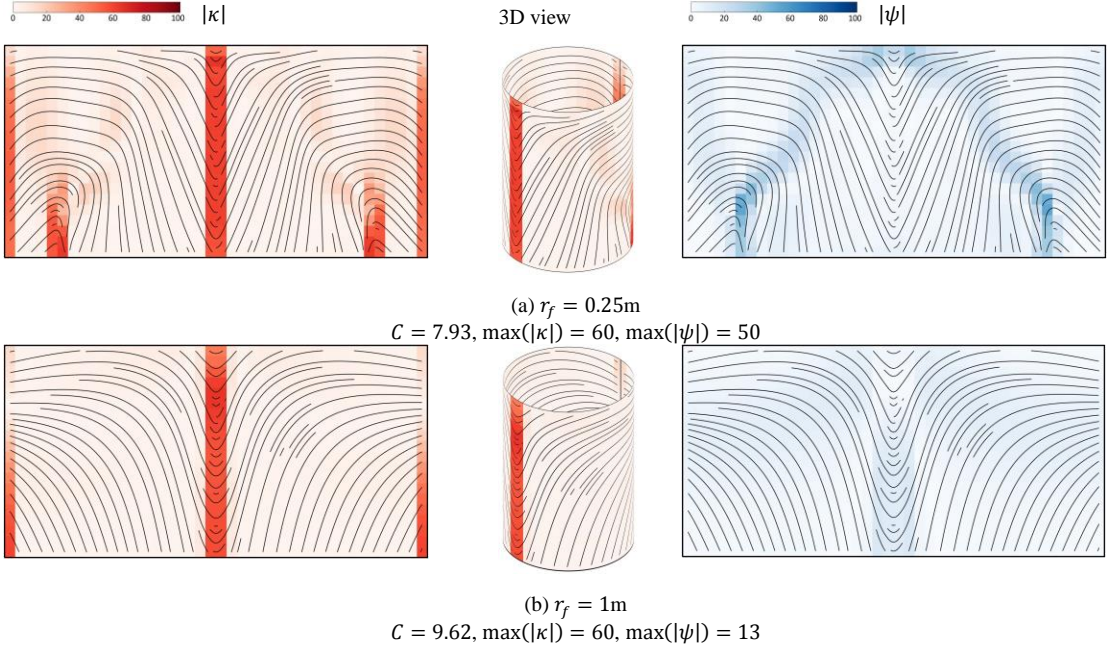

(a) $r_f = 0.25$m
$C = 7.93$, max($|\kappa|$) = 60, max($|\psi|$) = 50

(b) $r_f = 1$m
$C = 9.62$, max($|\kappa|$) = 60, max($|\psi|$) = 13

Fig. 13. Optimized designs for the single-ply cylindrical shell, obtained without manufacturing constraints and for two radii of the orientations filter. The left and right columns depict the orientation field on the 2D planar reference configuration, while the middle column provides a 3D perspective view.

Three designs obtained when including both curvature and divergence constraints, and for the orientation filter size $r_f = 0.25$m, are presented in Fig. 14. Setting the manufacturing constraints to $\overline{\kappa} = 50$ (1/m), $\overline{\psi} = 10$ (1/m) (thus restricting the maximum values attained by the design of Fig. 13(b)), some tow cut/add lengths are present near the middle line, suggesting the formation of gaps/overlaps (see Fig. 14(a)). Additionally, the tows near the middle and cutting lines still show a very high curvature. However, we can already see that the compliance is 8.5% lower than for the design in Fig. 13(b). The design in Fig. 14(b) employs more restrictive curvature constraints, resulting in flattened fiber paths with reduced curvature. Further reduction in gaps/overlaps is achieved by the design in Fig. 14(c) by applying stricter divergence constraints, leading to tow paths that align more closely with their neighbors.

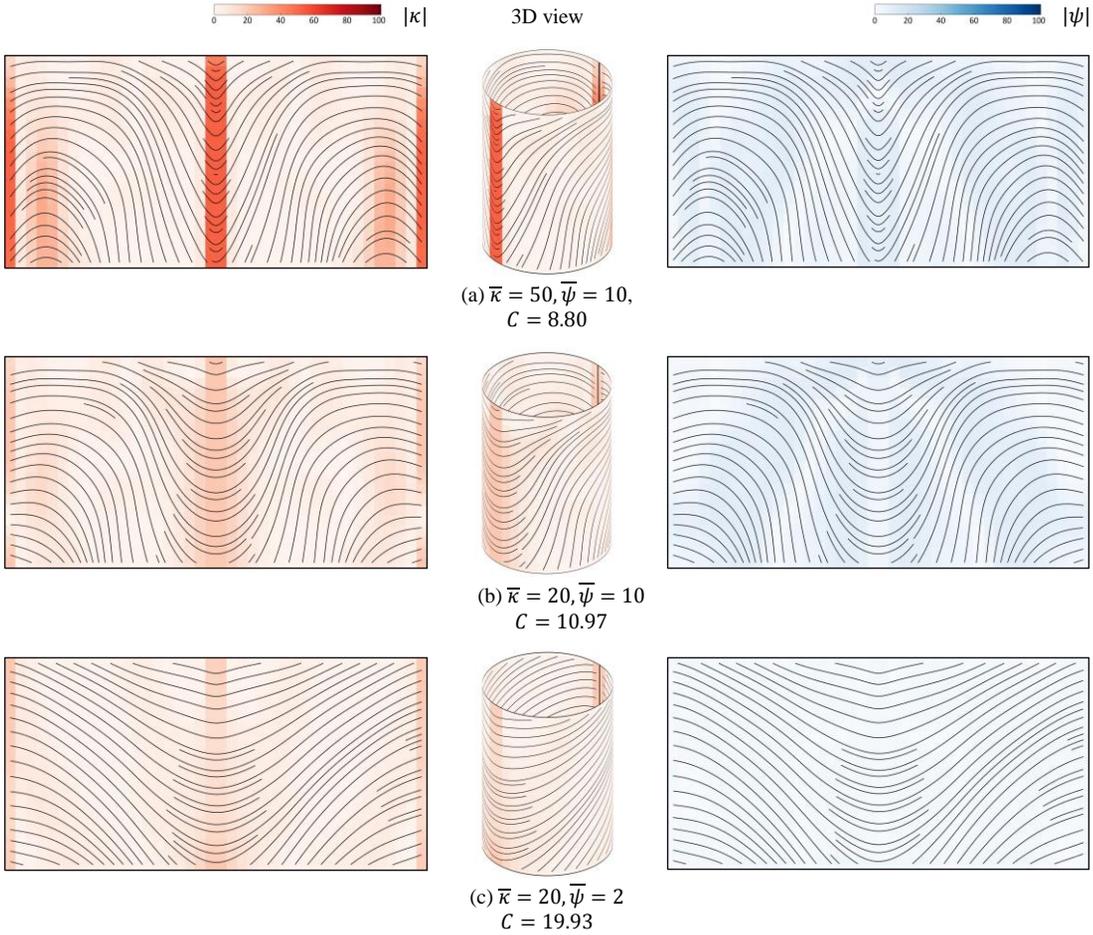

Fig. 14. Optimized designs for single-ply cylindrical shell with both curvature constraints and divergence constraints. In all cases, the manufacturing constraints are active and the constraint violation for the final design is less than 0.5%.

*Multi-ply design*

The proposed method is also tested on a four-ply composite laminated shell structure. The total thickness of the laminated structure is equivalent to the single-ply shell example. The initial design features a stacking sequence with uniform orientations [0°, 90°, 90°, 0°], with a compliance $C(s,t) = 18.41$J.

Fig. 15 shows the optimized solutions obtained applying both curvature and divergence constraints, with values $\bar{\kappa} = 20$ (1/m) and $\bar{\psi} = 1$ (1/m). The fiber path patterns are shown in both 2D and 3D for each layer, from the innermost (#1) to outermost (#4). Each of these layers has distinct tow orientations: layer #1 features more curved tows in the middle and in the cut line, while fiber path gradually inclines upwards from the middle to the cut line; layer #2 and layer #3 have near-vertical tows that lean slightly to the right and left, respectively; layer #4 shows more curved tows in the middle and in the cut line, while fiber path gradually inclines downwards from the middle to the cut line. The orientation fields on each ply are nearly free of gaps/overlaps, indicating excellent manufacturability with the tow steering process.

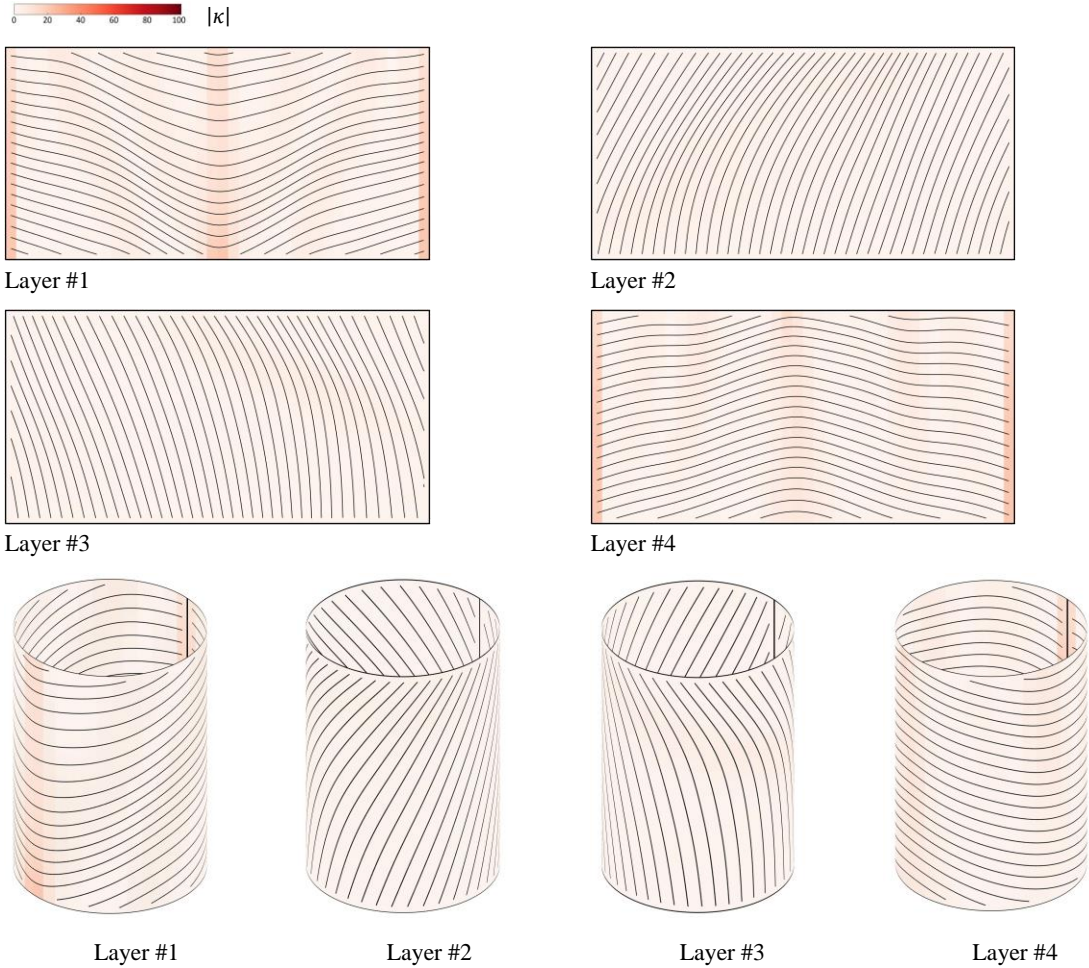

Fig. 15. Optimized design for multi-ply cylindrical shell under manufacturing constraint on the maximum curvature ($\overline{\kappa} = 20\ (1/m)$), and maximum density of gaps/overlaps ($\overline{\psi} = 1\ (1/m)$).

Fig. 16 displays the convergence history for the four-ply design. The optimized design exhibits a 46% reduction in structural compliance, compared to the initial design, with the compliance rapidly decreasing in the first 100 iterations, then being nearly constant. While the manufacturing constraints experience initial violations within the first 50 iterations, these violations steadily decrease, ultimately converging to near-zero values in the final stages of the optimization process, with some minor acceptable oscillations.

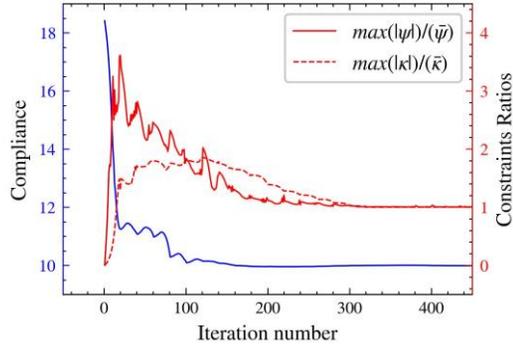

Fig. 16. Convergence plots for the four-ply cylindrical shell design: structural compliance (plotted against left axis), and manufacturing constraints ratios (plotted against the right axis).

## 6. Conclusion

We have proposed a method to incorporate important manufacturing requirements within the optimal design of variable stiffness laminates, featuring continuous fiber orientations. Specifically, we focus on imposing fiber continuity, reducing the maximum curvature of tows, and reduction of the gaps/overlaps densities within VSL composites. These requirements can be formulated in terms of the differential operators of the orientation field [36], and are imposed over each finite element in the design domain discretization. To efficiently manage the large number of resulting constraints, we implement an Augmented Lagrangian approach, which is shown to be effective and offer solutions that offer smaller compliance (larger stiffness) than found using other constraint aggregation strategies. The approach has been tested on minimum compliance design problems for 2D plates and 3D cylindrical shells, and for single-ply and multi-ply laminates. Results demonstrate that the proposed method can ensure that manufacturing constraints are locally fulfilled, allowing for precise control over tow orientation manufacturability.

Compared to conventional orientation filtering, the proposed method provides more precise control over the tow maximum curvature and maximum density of gaps/overlaps, while maintaining fiber continuity, and offers solutions with increased stiffness. We also note that a strict control of gaps/overlaps density seems to have a larger impact than curvature control, both in terms of compliance increase and in terms of the indirect control on the other manufacturing constraint. In all cases, relaxing manufacturing constraint magnitudes leads to designs with increased stiffness. This suggests that tow feedstock and manufacturing equipment that enable a smaller turning radius of steered tows and/or shorter cuts are one path towards enabling designs with larger stiffness.

## Acknowledgements

This work was supported by the National Aeronautics and Space Administration (NASA) Space Technology Research Institute (STRI) for Ultra-Strong Composites by Computational Design (US-COMP) (Grant No. NNX17AJ32G). Any opinions, findings, and conclusions or recommendations expressed in this article are those of the author(s) and do not necessarily reflect the views of NASA. The work of the second author was supported by the Villum Foundation, through the Villum Investigator project "InnoTop".

## Appendix A. 3D constitutive model

Given the strain vector $\boldsymbol{\varepsilon_0} = \{\varepsilon_{11}, \varepsilon_{22}, \varepsilon_{33}, \gamma_{12}, \gamma_{23}, \gamma_{13}\}$ and stress vector $\boldsymbol{\sigma_0} = \{\sigma_{11}, \sigma_{22}, \sigma_{33}, \sigma_{12}, \sigma_{23}, \sigma_{13}\}$ in material coordinates, the general orthotropic constitutive tensor, used in the 3D shell example of Section 5.3 is

$$\boldsymbol{C_0} = \begin{bmatrix} E_1(1-\nu_{23}\nu_{32})/\Delta & E_1(\nu_{21}+\nu_{23}\nu_{31})/\Delta & E_1(\nu_{31}+\nu_{21}\nu_{32})/\Delta & 0 & 0 & 0 \\ E_2(\nu_{12}+\nu_{13}\nu_{32})/\Delta & E_2(1-\nu_{13}\nu_{31})/\Delta & E_2(\nu_{32}+\nu_{12}\nu_{31})/\Delta & 0 & 0 & 0 \\ E_3(\nu_{13}+\nu_{12}\nu_{23})/\Delta & E_3(\nu_{23}+\nu_{13}\nu_{21})/\Delta & E_3(1-\nu_{12}\nu_{21})/\Delta & 0 & 0 & 0 \\ 0 & 0 & 0 & G_{12} & 0 & 0 \\ 0 & 0 & 0 & 0 & G_{13} & 0 \\ 0 & 0 & 0 & 0 & 0 & G_{23} \end{bmatrix} \quad (A.1)$$

where $\Delta = 1 - \nu_{12}\nu_{21} - \nu_{13}\nu_{31} - \nu_{23}\nu_{32} - \nu_{13}\nu_{32}\nu_{21} - \nu_{31}\nu_{12}\nu_{23}$. Since fibers lie in the plane of each ply, perpendicular to the shell thickness direction, the transformation matrix $\boldsymbol{Q}(m,n)$ is given by

$$\boldsymbol{Q}(m,n) = \begin{bmatrix} m^2 & n^2 & 0 & 2mn & 0 & 0 \\ n^2 & m^2 & 0 & -2mn & 0 & 0 \\ 0 & 0 & 1 & 0 & 0 & 0 \\ -mn & mn & 0 & m^2-n^2 & 0 & 0 \\ 0 & 0 & 0 & 0 & m & n \\ 0 & 0 & 0 & 0 & -n & m \end{bmatrix} \quad (A.2)$$

The cylindrical shell is discretized by a 3D shell element, which assumes zeros transverse normal stress $\sigma_{33}$ and is specifically developed for laminated composite shell structures (see details in [66,67]).

## Appendix B. Sensitivity analysis

The minimum compliance problem (11), (12) is self-adjoint, and the derivatives of the compliance $C(\boldsymbol{s}, \boldsymbol{t})$ with respect to the independent fiber orientation design variable $\boldsymbol{\vartheta} = (s, t)$ is simply computed by chain rule

$$\frac{\partial C}{\partial \boldsymbol{\vartheta}} = \frac{\partial C}{\partial \tilde{\boldsymbol{\vartheta}}} \frac{\partial \tilde{\boldsymbol{\vartheta}}}{\partial \bar{\boldsymbol{\vartheta}}} \frac{\partial \bar{\boldsymbol{\vartheta}}}{\partial \boldsymbol{\vartheta}} \quad (B.1)$$

The first term is found via the adjoint method and takes the well-known form of

$$\frac{\partial C}{\partial \tilde{\vartheta}^e} = -\boldsymbol{u}_e^T \frac{\partial \boldsymbol{K}^e}{\partial \tilde{\vartheta}^e} \boldsymbol{u}_e \quad (B.2)$$

where $\boldsymbol{u}_e$ is the vector of displacements associated with the element and $\boldsymbol{K}^e$ is the elemental stiffness matrix. The second term in Eq. (B.1) is computed by taking the derivative of Eq. (5), for example,

$$\frac{\partial \tilde{m}}{\partial \bar{s}} = \frac{\bar{t}^2 + \epsilon}{(\bar{s}^2 + \bar{t}^2 + \epsilon)^{3/2}}, \frac{\partial \tilde{m}}{\partial \bar{t}} = \frac{-\bar{s}\bar{t}}{(\bar{s}^2 + \bar{t}^2 + \epsilon)^{3/2}} \quad (B.3)$$

and the third term in Eq. (B.1) is simply

$$\frac{\partial \bar{\vartheta}^e}{\partial \vartheta^i} = \frac{w(X_i - X_e)}{\sum_i w(X_i - X_e)} \quad (B.4)$$

Similarly, the derivative of constraints in the AL function is obtained by chain rule as

$$\frac{\partial g}{\partial \boldsymbol{\vartheta}} = \frac{\partial g}{\partial \tilde{\boldsymbol{\vartheta}}} \frac{\partial \tilde{\boldsymbol{\vartheta}}}{\partial \bar{\boldsymbol{\vartheta}}} \frac{\partial \bar{\boldsymbol{\vartheta}}}{\partial \boldsymbol{\vartheta}} \quad (B.5)$$

Computing curvature and divergence constraints relies on spatial gradient information of the orientation field, which are computed by a finite difference scheme. Central finite difference is used to approximate Eq. (7) and (8) at interior data points, as shown in Fig. B.1(a)

$$\kappa^e = \frac{\tilde{n}_{e_{i+1,j}} - \tilde{n}_{e_{i-1,j}}}{2h_x} - \frac{\tilde{m}_{e_{i,j+1}} - \tilde{m}_{e_{i,j-1}}}{2h_y} \tag{B.6}$$

$$\psi^e = \frac{\tilde{m}_{e_{i+1,j}} - \tilde{m}_{e_{i-1,j}}}{2h_x} + \frac{\tilde{n}_{e_{i,j+1}} - \tilde{n}_{e_{i,j-1}}}{2h_y} \tag{B.7}$$

where $e_{i-1,j}$, $e_{i,j-1}$, $e_{i+1,j}$, and $e_{i,j+1}$ are the neighboring elements surrounding the element $e_{i,j}$. $h_x$ and $h_y$ are the size of finite elements. The corresponding derivative is

$$\frac{\partial \kappa^e}{\partial \tilde{m}_{e_{i,j+1}}} = -\frac{1}{2h_y}, \frac{\partial \kappa^e}{\partial \tilde{n}_{e_{i+1,j}}} = \frac{1}{2h_x}, \frac{\partial \kappa^e}{\partial \tilde{m}_{e_{i,j-1}}} = \frac{1}{2h_y}, \frac{\partial \kappa^e}{\partial \tilde{n}_{e_{i-1,j}}} = -\frac{1}{2h_x} \tag{B.8}$$

$$\frac{\partial \psi^e}{\partial \tilde{n}_{e_{i,j+1}}} = \frac{1}{2h_y}, \frac{\partial \psi^e}{\partial \tilde{m}_{e_{i+1,j}}} = \frac{1}{2h_x}, \frac{\partial \psi^e}{\partial \tilde{n}_{e_{i,j-1}}} = -\frac{1}{2h_y}, \frac{\partial \psi^e}{\partial \tilde{m}_{e_{i-1,j}}} = -\frac{1}{2h_x} \tag{B.9}$$

For data points at the edges or corners (see Fig. B.1(b) and (c) for example), a single-sided difference scheme is used.

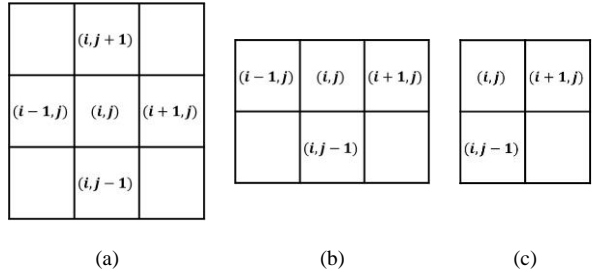

(a)    (b)    (c)

Fig. B.1. Neighboring elements involved in the finite difference scheme for element $e_{i,j}$ at different locations: (a) interior element, (b) a boundary element, and (c) a corner element.